\documentclass[letter]{aa} 
\usepackage{graphicx}
\usepackage{txfonts}
\usepackage{natbib}
\usepackage{xspace}
\usepackage{amssymb}
\usepackage{mathrsfs}
\bibpunct{(}{)}{;}{a}{}{,}
\def\deg{$^\circ$\xspace}
\begin{document}
%
   \title{On the interplay between flaring and shadowing in disks
     around Herbig Ae/Be stars.}

   \author{B. Acke
          \inst{1}\fnmsep\thanks{Postdoctoral Fellow of the Fund for
          Scientific Research, Flanders.}
         \and
          M. Min\inst{2}
          \and
          M.E. van den Ancker\inst{3}
          \and
          J. Bouwman\inst{4}
          \and
          B. Ochsendorf\inst{2}
          \and
          A. Juhasz\inst{4}
          \and
          L.B.F.M. Waters\inst{1,2}
          }

   \institute{Instituut voor Sterrenkunde, K.U.Leuven, Celestijnenlaan
              200D, B-3001 Leuven, Belgium\\
              \email{bram@ster.kuleuven.be}
         \and
             Sterrenkundig Instituut ``Anton Pannekoek'', Universiteit
             van Amsterdam, Kruislaan 403, 1098 SJ Amsterdam, the
             Netherlands 
         \and
             European Southern Observatory, Karl-Schwarzschild-Strasse 2,
             D-85748 Garching bei M{\"u}nchen, Germany 
         \and
             Max-Planck-Institut f{\"u}r Astronomie, K{\"o}nigstuhl 17,
             D-69117 Heidelberg, Germany
             }

   \date{Received --; accepted --}

 
  \abstract
   {Based on their spectral energy distribution, Herbig stars have been
categorized into two observational groups, reflecting their overall
disk structure: group~I members have disks with a high degree of flaring as
opposed to their group~II counterparts. Literature results show that
the structure of the disk is a strong function of the disk mass in
$\mu$m-sized dust grains.}
   {We investigate the 5--35~$\mu$m Spitzer IRS spectra of a sample of
13 group~I sources and 20 group~II sources. We focus on the continuum
emission to study the underlying disk geometry.} 
   {We have determined the [30/13.5]
and [13.5/7] continuum flux ratios. The 7-$\mu$m flux excess with
respect to the stellar photosphere is measured, as a marker for the
strength of the  
near-IR emission produced by the hot inner disk. We have compared our
data to the spectra produced by self-consistent passive-disk models,
for which the same quantities were derived.}
   {We confirm the results by Meijer et al. (2008) that the
differences in continuum emission between group~I and II sources can
largely be explained by a difference in amount of small dust
grains. However, we report a strong correlation between the [30/13.5]
and [13.5/7] flux ratios for Meeus group~II sources. Moreover, the
[30/13.5] flux ratio decreases with increasing 7-$\mu$m excess for all
targets in the sample. To explain these correlations with the models,
we need to introduce an artificial scaling factor for the inner disk
height. In roughly 50\% of the Herbig Ae/Be stars in 
our sample, the inner disk must be inflated by a factor 2 to 3 beyond
what hydrostatic calculations predict.} 
   {The total disk mass in small dust grains determines the degree of
     flaring. We conclude, however, that for any 
given disk mass in small dust grains, the shadowing of the
outer (tens of AU) disk is determined by the scale height of the inner
disk ($\sim$1~AU). The inner disk partially obscures the outer disk,
reducing the disk surface temperature. Here, for the first time, we
prove these effects observationally.}


   \titlerunning{Flaring and shadowing in Herbig Ae/Be disks.}
   \authorrunning{B. Acke et al.}

   \maketitle
%

\section{Introduction}

Herbig Ae/Be stars are the intermediate-mass (2 -- 8 M$_\odot$)
analogues of T~Tauri stars. They are surrounded by circumstellar
disks, which are believed to be the sites of ongoing planet
formation. The geometry of these disks has been the subject of
numerous studies. \citet{meeus01} have classified the 
observed spectral energy distributions (SEDs) of a sample of Herbig
stars into two groups: group~I members display a significantly larger
far-infrared excess than their group~II counterparts. \citet{dullemond04}
have provided a physical explanation for this difference: group~II
sources have an outer disk which is protected against direct stellar
radiation by the puffed-up inner disk. However, if the outer disk
emerges from the inner disk's shadow, i.e. has a large flaring angle,
its SED resembles that of a group~I source. \citet{dullemond04b} and
\citet{meijer08} indicate that the dust grain size distribution plays 
a determining role in the structure of these circumstellar disks and
their SED appearance. This idea is supported by observational
evidence: the grains responsible for the 
(sub-)millimeter emission are significantly smaller in group I than in
group II \citep{ackesubmm}. 

In the early disk models, the inner disk was modeled as a vertical
{\em wall} at the dust sublimation distance to the star. Later
refinements \citep[e.g.,][]{isella05} took into account the dependence of grain
sublimation temperature on gas density, resulting in a curved inner
rim. With the coming of age of near-IR interferometers, however, it
became clear that our knowledge of the inner disk remains
incomplete. For a growing list of targets
\citep{monnier05,kraus08,acke08,tannirkulam08}, the inner disk appears 
less spatially resolved than expected from the models. This is
commonly explained by the presence of an optically thick, possibly
gaseous, disk component located within the dust sublimation radius.
However, near-IR interferometric data have been successfully fitted
with artificially increased inner disk heights as well (Verhoeff et
al. 2009, submitted to A\&A). 


We have collected and analyzed the Spitzer IRS \citep{houck04} spectra
of a sample of 54 Herbig Ae/Be stars (Bouwman, Juhasz, Acke et al. in
prep).
In this Letter, we focus on the shape of the underlying continuum and
report the discovery of two tight correlations
between the near-IR excess and mid-to-far-IR colors. We explain these
correlations in the framework of self-consistent passive-disk models
\citep{min09}.


\section{Anti-correlation between near-IR excess and far-IR color}

\begin{figure}
\centering
\includegraphics[width=\columnwidth]{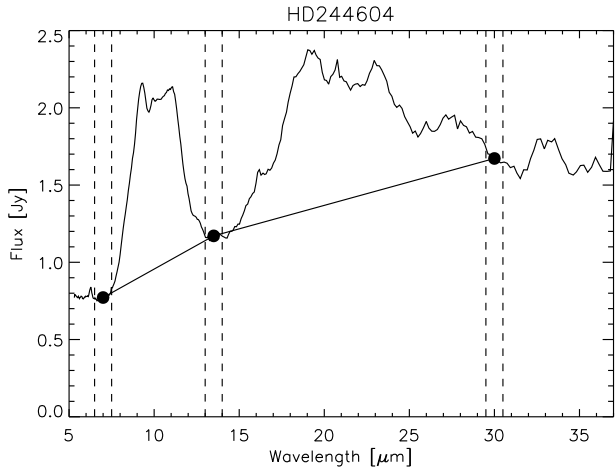}
\caption{Spitzer IRS spectrum of the sample target HD244604. Indicated
  are the
  continuum wavelengths at 7, 13.5 and 30~$\mu$m where 
  the spectral features of silicates and PAHs are weak or absent. The
  bin width is 1~$\mu$m.}
           \label{IRcolors_method.ps}%
 \end{figure}

\begin{figure*}
\centering
\includegraphics[width=0.93\textwidth]{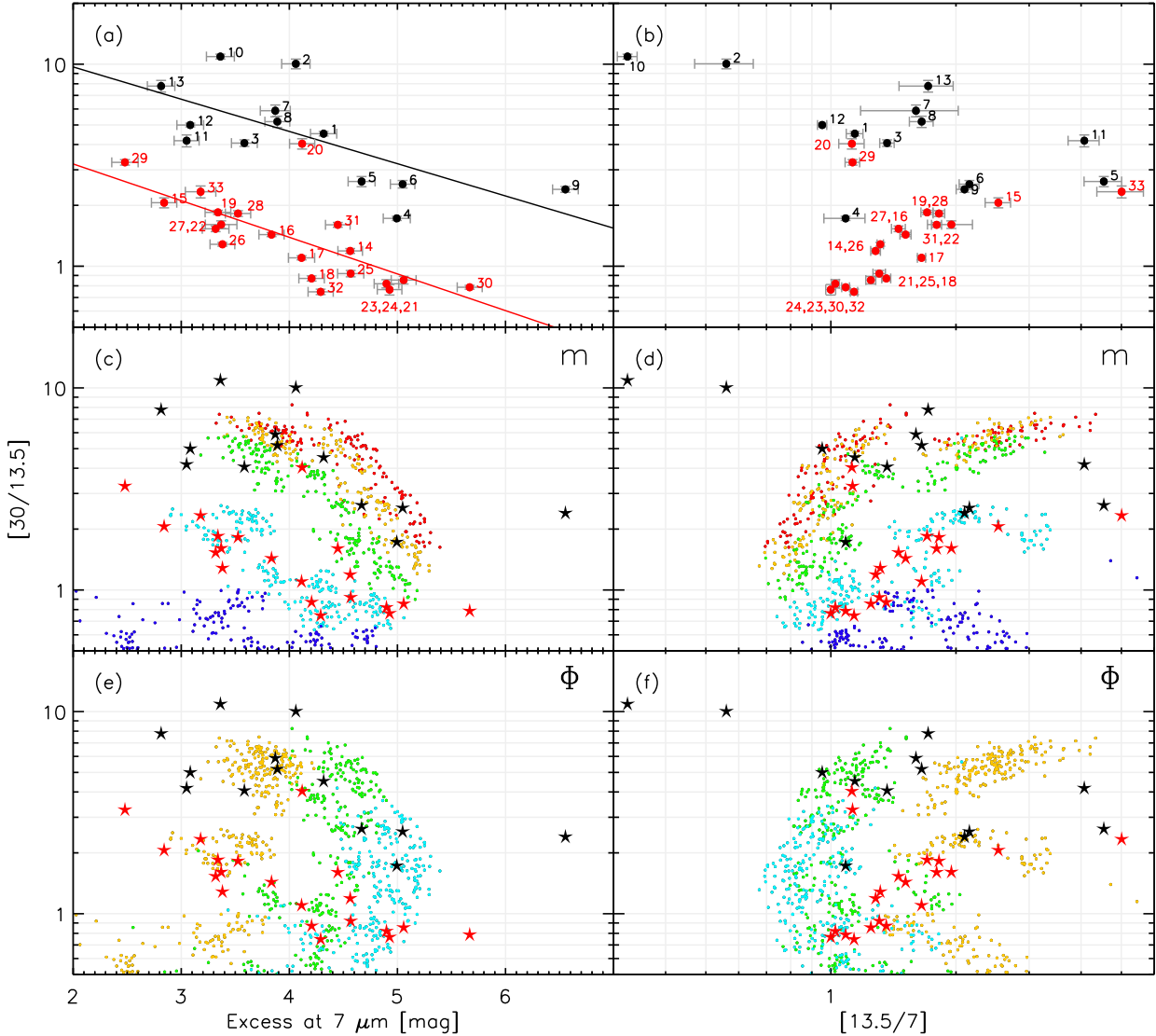}
\caption{{\em Left panels: (a)} The clear anti-correlation between the
7-$\mu$m excess and the Spitzer [30/13.5] flux ratio. Meeus group~I
and II sources are resp. black and red dots. Also shown are the best
linear fits to the correlation for group~I (black) and II (red line).
Group~I: 1. \object{HD31293}, 2. \object{HD34282},
3. \object{HD36112}, 4. \object{RR Tau}, 5. \object{HD38120},
6. \object{HD250550}, 7. \object{HD97048}, 8. \object{HD100453}, 9.
\object{SS73 44}, 10. \object{HD135344B}, 11. \object{HD139614},
12. \object{HD142527}, 13. \object{HD169142}, Group~II:
14. \object{HD31648}, 15. \object{HD35187}, 16. \object{HD244604},
17. \object{HD37258}, 18. \object{BF Ori}, 19. \object{HD37357},
20. \object{HD37411}, 21.  \object{HD37806}, 22. \object{HD72106S},
23. \object{HD85567}, 24. \object{HD95881}, 25.  \object{HD101412},
26. \object{HD104237}, 27. \object{HD142666}, 28. \object{HD144432},
29.  \object{HD152404}, 30. \object{VV Ser}, 31. \object{WW Vul},
32. \object{HD190073}, 33.  \object{HD203024}. {\em (c) and (e):}
Identical to panel (a) with all models included (dots). The stars
refer to the observations. (c) The color of the dots refers to the
total disk mass in small dust grains: $\log m$ [M$_\odot$] = $-$3 to
$-$7 from red to blue. The models shift from upper right to lower left
with decreasing mass in small dust grains. (e) Colors refer to the
inner disk scale height parameter $\Phi$: yellow are the models in
hydrostatic equilibrium, green and blue refer to $\Phi$ equal to 2 and
3. The models shift from upper left to lower right with increasing
inner rim scale height. {\em Right panels: (b)} The [30/13.5] flux
ratio as a function of the [13.5/7] ratio. A tight correlation between
both flux ratios is noted for group II. {\em (d) and (f):} Identical
to panel (b) with models included (dots). The same plotting convention
as in the left panels is used. The models shift from upper left to
lower right with decreasing mass in small dust grains (d) and from
upper right to lower left with increasing scale height (f).}
           \label{IRcolors_BIG.ps}%
 \end{figure*}

For all the target stars, we have fitted a reddened \citet{kurucz}
model to optical photometric measurements available in the literature. 
From the IRAS~60~$\mu$m photometry, the excess magnitude is computed
(M$_{\mathrm{exc}}$ = 2.5 $\log$(F$_{\mathrm{tot}}$/F$_{\star}$)). 
The Spitzer IRS spectra are used to determine the average fluxes at 7,
13.5 and 30~$\mu$m in a bin with a width of 1~$\mu$m. As shown
in Fig.~\ref{IRcolors_method.ps}, only a few wavelengths are suitably
free of solid-state features. The chosen bin width is a compromise
between maximal signal-to-noise and minimal influence of spectral
features of amorphous and crystalline silicates, and PAHs. The error
bars on the measured fluxes 
include the errors in the individual spectral channels and the
spread of the spectrum within the bin. 
In this paper we discuss the [13.5/7] and [30/13.5] flux ratios, and
the 7-$\mu$m excess which probes the near-IR excess\footnote{Approx.
70\% of the 7-$\mu$m flux is produced in the innermost disk}. The error on 
the latter includes an additional 0.11~mag, as the photometric
accuracy of Spitzer IRS is 10\% \citep{bouwman08}. 

In previous work \citep{ackenano}, we have shown that the IRAS 60~$\mu$m
excess can distinguish in a simple and reliable way between the two
\citet{meeus01} groups. Targets with a 60 $\mu$m excess
exceeding 10.1 magnitudes are classified as group~I, the others as
group~II. Seven sources with tenuous or optically thin disks, defined as
those with a total infrared-to-stellar luminosity
L$_\mathrm{IR}$/L$_\star$ below 10\%, are excluded as they 
cannot be compared to our optically thick disk models
(Sect.~\ref{models}). Our final sample contains 13 group~I and 20
group~II sources for which Spitzer IRS spectra are available covering
the full range from 5 to 35~$\mu$m.

In Fig.~\ref{IRcolors_BIG.ps}{\em a}, the [30/13.5] flux ratio is
plotted as a function of the excess at 7~$\mu$m.  For both group~I and
group~II sources, a clear decline in [30/13.5] flux ratio is observed
with increasing 7-$\mu$m excess. As indicated by the Pearson
correlation coefficient, the trend is significant with a
false-positive probability (FPP) less than 1\%. We have performed a
linear fit to the (7-$\mu$m excess, $\log [30/13.5]$) data
of group~I and II separately. The fits, indicated by the black
and red line in the figure, have similar slopes: $-0.16 \pm 0.01$ and
$-0.18 \pm 0.01$ for group~I and II respectively. Note, however, that
the [30/13.5] flux ratios in group~I are higher than those in group~II
for any given value of the 7-$\mu$m excess.

Fig.~\ref{IRcolors_BIG.ps}{\em b} is a plot of the [30/13.5]
versus [13.5/7] flux ratio. Group II sources display a pronounced
correlation between both colors (FPP $<$ 1\%). The
group~I sources are scattered in this diagram. A few outliers are
noted in the plot. Group~I sources HD34282 (\#2) 
and HD135344B (\#10) are the bluest sample sources in [13.5/7] color,
and the reddest in [30/13.5]. These targets have indications for the
presence of a large (opacity) gap in their inner disk, which explains
their anomalous colors. HD37411 (\#20) and HD152404 (AK~Sco, \#29)
have the highest [30/13.5] ratio in group~II and deviate from the
trend. Also for these sources, a large disk gap may be present. The
presence of a (stellar or sub-stellar) binary companion likely plays a
role in opening up the gap.

\section{Comparison with disk models \label{models}}

\begin{table}
\begin{minipage}[t]{\columnwidth}
\caption{ Model parameter ranges. $m$ is the mass
  in sub-$\mu$m-sized dust grains, $p$ is the power which
  governs the decline of the surface density with increasing distance
  to the star ($\Sigma \propto R^{-p}$), $i$ is the inclination of the
disk with respect to the line-of-sight, $\Phi$ is the artificial
scaling factor for the inner disk scale height, with $\Phi = 1$
indicating hydrostatic-equilibrium models. See text for
details.}
\label{tab_modelpar} 
\renewcommand{\footnoterule}{}  
\centering   
\begin{tabular}{ll} 
\hline\hline    
Parameter & Values \\
\hline
$\log m$ [M$_\odot$ ] & $-3, -4, -5, -6, -7$\\  
$p$     & 1, 1.5, 2 \\
$i$     & 10\deg $-$ 90\deg \\
$\Phi$  & 1, 2, 3 \\
\end{tabular}
\end{minipage}
\end{table}

From the correlations described above, it appears that the hot inner
disk ($\sim$1 AU) and the structure of the outer disk (tens of
AU) closely relate to each other in the sample of Herbig stars. To
understand 
the physics behind this result, we have computed a grid of
disk models using the radiative transfer code MCMax \citep{min09}. The
central star in these models is a main-sequence star of
spectral type A0 (T$_\mathrm{eff}$ = 10\,000\,K, R$_\star$ = 2 R$_\odot$,
M$_\star$ = 2.5 M$_\odot$), representative for the median star in the
sample. Altering the stellar parameters over the sample range has no
influence on the analysis below.

A circumstellar disk was modeled, assuming hydrostatic equilibrium
(HSE) and thermal coupling between gas and dust. The gas-to-dust
mass ratio was fixed to 100. The dust consists of astronomical
silicates with a grain size of 0.12~$\mu$m. Large ($>$ several
$\mu$m) grains have not been included, as they do not influence the
SED shape shortward of 60~$\mu$m \citep{meijer08}. The
total dust mass in the models ($m$) hence only refers to the total
dust mass in these small grains and should be interpreted as such. We
have computed models with different values for $m$.
Next to the dust mass, the surface density power law ($\Sigma
\propto R^{-p}$) and the inclination of the system ($i$) were
altered. Also, we have 
artificially increased the inner disk's scale height by a factor
$\Phi$, to study the effect of a more puffed-up inner rim than HSE
calculations produce. The ranges of the model parameters are listed in
Table~\ref{tab_modelpar}. 
As for the sample targets, only disk models that display a
10~$\mu$m silicate {\em emission} feature in their spectrum are
included in the 
analysis. In total, 1000 disk models were included in the study.
Based on their 60 $\mu$m excess, roughly half would be 
classified as Meeus group~I sources, the other half as group~II.

Varying the inclination of the considered disk models only has a minor 
influence on the 7-$\mu$m excess (variations of 0.2~mag) and the IR
flux ratios (25\%). A spread in inclinations
therefore cannot reproduce the range, location and correlations of the
observations in the diagrams.
The locus of the models is predominantly determined by
the mass in small dust grains (see Fig.~\ref{IRcolors_BIG.ps}{\em
  c,d}). The higher the mass in small dust grains, the 
larger the outer disk's flaring angle becomes, and hence the redder
the model appears in the [30/13.5] flux ratio. However, the dust mass
cannot account for the correlation that is seen between the 7-$\mu$m
excess and the [30/13.5] flux ratio in both groups, and the
correlation between the [13.5/7] and [30/13.5] flux ratios in
group~II.

The power of the surface density law has a moderate influence on
the location of the models
in the diagrams: models with a steep decline of the surface density
with radius are bluer. However, the changes induced by this parameter
are insufficient to explain the observed spread in disk excess and
color. To cover the full observational range, especially in 7-$\mu$m
excess, we need to artificially inflate the 
inner disk beyond its equilibrium scale height by a factor of a
few (see Fig.~\ref{IRcolors_BIG.ps}{\em e,f}). 

\begin{figure}
\centering
\includegraphics[width=\columnwidth]{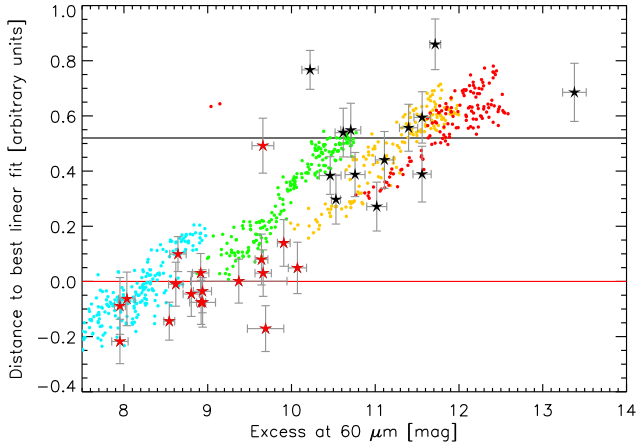}
\caption{The distance to the best linear fit to the
  group~II data in Fig.~\ref{IRcolors_BIG.ps}{\em a} (red line) is
  strongly correlated with the 60-$\mu$m excess. The
  black line represents the best fit to the group~I data in
  the same figure. The stars are the group~I (black) and group~II
  data (red), the dots are the models. Colors reflect the mass in
  small dust grains, from $m = 10^{-6}$ (bottom left) to $10^{-3}$
  M$_\odot$ (top right).}
           \label{IRcolors_distvsIRAS60.ps}%
 \end{figure}

The correlation between the [30/13.5] color and the 7-$\mu$m excess
displays scatter that greatly exceeds the observational errors.
We take the best linear fit to the group~II data (i.e. the red line in
Fig.~\ref{IRcolors_BIG.ps}{\em a}) as the reference and compute the
perpendicular distance of all observations to this line. This distance
is strongly correlated with the magnitude of the 60~$\mu$m excess:
sources below the average trend have lower excesses than targets above
it (see Fig.~\ref{IRcolors_distvsIRAS60.ps}). This correlation with
far-IR excess is significant for the sample as a whole (FPP $<$ 1\%),
but also when limited to group~II sources alone (FPP $<$ 5\%).  The
disk models also display the increase in 60~$\mu$m excess with
distance to the average fit. This confirms that the mass in small dust
grains determines the locus of the targets and the disk models in the
color-excess diagram, {\em perpendicular} to the observed
correlation. Group~I disks are more massive in small dust grains ($m >
10^{-5}$ M$_\odot$), while the group~II disks cluster around $m =
10^{-6}-10^{-5}$ M$_\odot$. {\em Along} the color-excess and
color-color correlations, the defining parameter is the scale height
of the inner disk. Roughly half of the targets can be modeled with a
disk in hydrostatic equilibrium. The disks that are blue in [13.5/7]
and [30/13.5] colors, however, have large 7-$\mu$m excesses. For these
sources, an artificially increased scale height of the inner disk is
required in our models.

\section{Discussion}


The degree of flaring is the increase in opening 
angle of the disk surface $\alpha(r)$ (= $\arctan(h(r)/r)$, with
$h(r)$ the 
height of the surface above the midplane) with increasing distance to
the star $r$. The disk surface is the surface that
radiates in the infrared, where the opacity along the line-of-sight
reaches unity. All the model disks flare, but the flaring
is stronger in models with a high mass in small dust
grains ($h(r) \propto r^{1.30}$ vs. $r^{1.15}$). Moreover, the maximal
opening angle of a massive disk is approximately 5 times that of a
low-mass disk ($\alpha_\mathrm{max} = 25^\circ$ vs. 5$^\circ$).

The observed range in 7-$\mu$m excess surpasses the range predicted by
hydrostatic-equilibrium models. An inflation of the inner disk height,
beyond what HSE models produce, naturally leads to a higher near-IR
excess, but also affects the outer disk. A disk with a puffed-up
inner rim casts a large shadow over the 
outer disk. The surface layers of the outer disk are therefore cool
and will produce hardly any infrared excess in the mid-to-far-IR. The
combination of both effects reduces the [30/13.5] and [13.5/7] flux
ratios with increasing 7-$\mu$m excess and explains
the observed correlations. Note that the opening angle and degree of
flaring of the disk is set by the mass in small dust grains, and only
marginally depends on the height of the inner disk:
$\alpha_\mathrm{max}$ decreases by 10\% when $\Phi$ is increased from
1 to 3 for a given model. 

\section{Conclusions}

Using Spitzer IRS spectra of a large sample of Herbig Ae/Be stars
together with a large parameter set of radiative transfer models we
come to the following conclusions.

\begin{itemize}
\item There is a strong observational correlation between the
  near-infrared excess and the infrared spectral slope of Herbig Ae/Be
  stars. This indicates that the apparent geometry of the outer disk
  is determined by that of the inner disk.
\item Using radiative transfer modeling we show that the structure of
  the disk is governed by two parameters:
 \begin{enumerate}  
 \item The mass in small dust grains which determines the degree of
   flaring. 
 \item The inner disk scale height which determines the degree of
   shadowing.
 \end{enumerate}
\item Disk models in hydrostatic equilibrium fail to reproduce the
  observed spread in inner disk scale heights.
\end{itemize}

The findings presented in this letter together with literature
infrared interferometric results, show that our understanding of the
innermost regions of Herbig stars is yet incomplete. It is clear that
additional physical processes, not included in our models, play a
role. The correlation between the 7~$\mu$m excess and the apparent
geometry of the outer disk shows that the mechanism causing the flux
at short wavelengths influences the outer disk. This rules out
optically thin hot gas emission as an important contributor to the
7~$\mu$m excess in Herbig stars. Interestingly, recent modeling
efforts including gas chemistry calculations \citep[e.g.][]{woitke09}
show that the gas temperatures can be much higher than the dust
temperatures in the inner disk. A similar effect is obtained in disk
models when including the stellar far-UV to X-ray radiation field
\citep[e.g.][]{gorti09}. As a result the inner disk height is strongly
increased, while the outer disk remains unaltered. Further study is
needed to verify whether this affects the disk continuum emission in
conformance with the observations.

\acknowledgements{This work is based on observations made with
  the Spitzer Space Telescope, which is operated by the Jet Propulsion
  Laboratory, California Institute of Technology under a contract with
  NASA. BA acknowledges support from the ESO visitor programme during
  the preparation of this manuscript.}

\bibliographystyle{aa}
\bibliography{/lhome/bram/REFERENCES/references.bib}

\end{document}